\documentclass[a4paper,11pt]{article}

\usepackage{jheppub} 
\usepackage{amsmath,amssymb,amsfonts}
\usepackage{mathtools}
\usepackage{amsfonts}
\usepackage{microtype}
\usepackage{setspace}
\usepackage{lineno,hyperref}
\usepackage{array,multirow,graphicx}
\usepackage{float}
\usepackage[font=scriptsize]{caption}
\usepackage[T1]{fontenc} 

\title{General Relativistic Thick Disks in the Accelerating Expanding Universe Dominated By Dark Energy}

\author[a]{Senobar Doostali }
\author[a]{, Alireza Mirzaee }
\author[a]{, Reza Ramezani Arani }
\author[b,1]{, Salman Abarghouei Nejad \note{salmanabar@gmail.com.}}
\affiliation[a]{Department of Physics, University of Kashan, 87317-51167, Kashan, Iran.}
\affiliation[b]{Departamento de Física, Universidade Federal de Pernambuco, 50670-901, Recife, Braszil.}

\abstract{In this article, we study the effect of the universe accelerating expansion on the structure of the general relativistic thick disks. Hence, by applying a conformal transformation on the metric of a static disk in isotropic coordinates, we transform the disk to the spacetime of a Friedman-Robertson-Walker ($FRW$) of an expanding universe. Also, by solving Einstein equations in the presence of the cosmological constant $\Lambda$ which leads to the expansion of the universe, we investigate the structure of this disk in a dynamical state, and the changes in the disk mass density and energy density, as well as radial, azimuthal and vertical pressure, are examined as a function of time. Furthermore, by considering weak, strong and dominant energy conditions, we find that relativistic thick disks in the accelerating expanding universe dominated by dark energy satisfy all energy conditions.}

\keywords{General Relativistic Thick Disks, Accelerating Expanding Universe, Dark Energy, Conformal Transformation.}

\begin{document} 
\maketitle
\flushbottom

\section*{Introduction}

Solving Einstein equations for thick disks has become a topic of interest in recent years. Indeed, thick disks are good approximations for various physical objects. Hence, in an accelerating expanding universe\cite{acc1,acc2}, solutions of general relativistic disks may help us to understand the mainspring of the accelerating expansion.

The exact solution to the Einstein equations are often studied in ideal physics systems with special geometrical properties. Nonetheless, it is possible to obtain important results on the properties of large physical system using simple exact solutions under certain conditions. Additionally, the exact solutions with axial symmetry play a critical role in astronomical applications of general relativity because the natural form of an isolated self-gravitating fluid is in fact symmetric. In particular, disk-like formations of matter are of special interest due to the fact that they can be utilized as models of galaxies and integrated disk.

The solutions to thin disks without radial pressure were initially studied by Bonner and Sackfield \cite{bib1}, and Morgan and Morgan \cite{bib2}. Such solutions were also calculated by Morgan and Morgan for static thin disks in the presence of radial pressure \cite{bib3}. Additionally, various exact solution to Einstein equations were found for static thin disks with, or without, radial pressure between 1970 and 2004.

Thin rotating disks were studied in \cite{bib4} by Bicak and Ledvinka. A new method was invented through general relativity in galactic dynamics modeling by Cooperstock and Tieu in which galaxies were considered as a rotating axially symetric fluid without radial pressure\cite{bib5}. Several authors have shown that the distribution of the density of galaxies is consistent with that of matter's mass \cite{bib6}.
Although thin disks can be used as a model for galaxies in the first approximation, the thickness of the disks needs to be taken into account while examining more realistic models. Relativistic thick disks in different coordinates have been described in \cite{bib7}. 

In all the above-mentioned disks, an inverse method has been employed to solve Einstein equations: the metric of the disk is first assumed and then used in the calculation of the source (energy-momentum tensor). This method is referred to as the "g-method" which is the opposite of the "t-method", i.e. the direct method of using the source to solve Einstein equations. Before this, there were some solutions developed for thin disks.

In this article, we sought to find solutions for thick disks by using "displace, cut, fill and reflect" method, which is a more realistic model of galaxies based on thin disk's solutions. This method, first introduced by Vogt and Letelier in 2005 \cite{bib7}, is one of the "g-method" solutions, i.e. first the metric is guessed then the elements of energy-momentum tensor could be found by using Einstain's equations. This has been done in \cite{bib7} , \cite{bib8}, \cite{bib9} for which a summary is presented in section \ref{section2}. But, one should notice that, all aforementioned studies have focused on thin or thick disks in static spacetime. However, solving the problem in an expanding universe, although adding to the complexity, could offer a solution closer to reality to understand the leading mechanisms behind this physical phenomenon.

On the other hand, although there are many alternative explanations for the accelerating universe \cite{CM1}, \cite{CM2}, \cite{CM3}, \cite{CM4}, \cite{CM5}, \cite{CM6}, we want to solve Einstein equations in cylindrically symmetric coordinates in an accelerating expanding universe to capture a more realistic dynamic for the universe. Thus, we first find the solutions in static spacetime, and then by applying conformal transformations, the solutions in displace, cut, fill and reflect" method are obtained. Vaidya \cite{bib10} and Pattel and Trrivedi \cite{bib11} computed the appropriate metric for Kerr black holes in FRW spacetime by applying Kerr-Schild transformations to FRW spacetime. The effect of acceleration is considered by adding cosmological constant in Einstein equations and the cosmological constant depends on the equation of state which is related to dark energy.

In this article, we first utilize "displace, cut, fill and reflect" method to make static thick disk, then we apply conformal transformation to calculate the metric of the disk in the spacetime of an expanding FRW universe. Lastly, by considering the effect of acceleration of universe expansion, solve Einstein equations in the present of the cosmological constant for a relativistic thick disk. It is worth noting that we use latest experimental data to estimate the cosmological constant.

The Outline of the article is as follows: in section \ref{section2}  a brief explanation of the "displace, cut, fill and reflect" method is described. Section \ref{section3} describes the mechanism we used to transform the metric in the static spacetime into a metric in expanding spacetime by applying the conformal transformation. Particularly, in the subsection \ref{section3.2}, by utilizing conformal transformations on a static thick disk we obtain the solution of thick disks in expanding spacetime. Finally, considering the effect of cosmological constant, we express the effect of dark energy on the acceleration of the expanding the universe and find scale factor $a(t)$ in \ref{section3.3} , and numerical solutions of the problem will be calculated and presented in \ref{section3.4}.

\newpage
\section{Constructing thick disks using ``displace, cut, fill and reflect" method}\label{section2}
\subsection{Introducing the ``displace, cut, fill and reflect" method}\label{section2.1}
Using the "displace, cut, and reflect" method, we can solve a thin disk by solving a Schwarzschild point source. Here, in the first step, we choose a plane that divides the space into two parts: a part without singularities or sources, and a part with singularities. In the second step, the part of the space with singularities will be neglected. In the third step, we will use a surface to build the inverse of the non-singular part of the space. The result will be a space with a singularity of the form of delta function with support on $z = 0$ plane. 

To generate a thick disk, we need to expand the above-mentioned procedure. Essentially, we need to replace the surface of discontinuity with a thick shell such that the matter comprising the disk can be described with continuous functions. In this case, our method consists of an additional step, which will be henceforth referred to as "displace, cut, fill and reflect". After ignoring the part of the space containing singularity, we will place a thick shell underneath the plane and then we will use the lower surface of this shell to build the inversion.
This implies that instead of plotting the initial space and the quarantined space z=0, by applying $z \to h(z) + b$ transform, the initial space is converted to $z=a$ and the space mapped to $z=-a$ and bound together.

Here, $-a<z<a$ is a region in which the source density exists and is related to the internal solution. Therefore, in $z>a$ , $z<-a$ the external solution of a thick disk is obtained. The amount or type of h(z) depends on the density and the stress of the disk. This can be obtained by guessing the metric to the energy density and disk tensions and can also be obtained by knowing the density and tensions of the disk to find the metric.

In the current manuscript, we start with the source of the Schwarzschild point and in fact with the above method we assume that the Schwarzschild solution of the external solution in $z>a$ , $z<-a$ is equivalent to solving a thick disk then we find the source density and disk tension which are the momentum-energy tensor elements. The point is to apply a decent h(z) to get the Schwazschild solution closer to solving a thick disk.

A subset of exact solutions to Einstein's field equations were proposed by Van Stockum \cite{bib12} to describe particle distributions with stable axial symmetry. These solutions have the following line element,
\begin{equation} \label{eq1}
 d{s^2} = {(dt - Nd\varphi )^2} - {r^2}d{\varphi ^2} - {e^\nu }(d{r^2} + d{z^2}),
\end{equation}
and the Einstein equations will be as ${\Phi _{,rr}} + {\Phi _{,zz}} + \frac{\Phi _{,r}}{r} = 0$ \footnote{$  \Phi_i$ is  the gravitational potential on a linearized scenario.}. To obtain a thick disk, considering the solutions to equation \eqref{eq1} , we apply the transformation $z \to h(z) + b$ where $b$ is a positive constant and $h(z)$ is an even function of $z$ and $h(0) = 0$. The parameter $b$ controls the amount of matter which needs to be concentrated in the proximity of the symmetry axis. Large amount of $b$ generates a more fluid mass distribution, such that by increasing the value of $b$, the rate of mass distribution from the center to the edge of the disk will decrease. Further, the distribution of the function $h(z)$ forces the mass distribution to be along the $z$ axis. This method is equivalent to constructing a disk using the "displace, cut, fill and reflect" method.

\subsection{A static solution of thick disks from Schwartzschild metric in conformal coordinates}\label{section2.2}
	At first we review the static solution of thick disks which studied before. As shown in \cite{bib7}, in the static spacetime and in the isotropic cylindrical coordinates, the metric of a thick disk from the Schwarzschild metric can be found.\\
In cylindrical coordinates $(t, r, z, \phi)$, the conformal metric is as follows,
\begin{equation} \label{eq2}
\ d{s^2} = - {e^{2\nu }}d{t^2} + {e^{2\mu }}(d{r^2} + d{z^2} + {r^2}d{\varphi ^2}),
\end{equation}
where $\mu$ and $\nu$ are functions of $r$ and $z$.

The Schwarzschild solution for the metric in equation \eqref{eq2} are described as,
\begin{equation}\label{eq3}
\begin{array}{l}
\nu = ln[\frac{{2R - m}}{{2R + m}}] \\
\mu = 2ln[1 + \frac{m}{{2R}}]
\end{array}
\end{equation}
where $m>0$ and ${R^2} = {r^2} + {z^2}$. By applying the ``displace, cut, fill and reflect" method on the above solutions we find, ${R^2} = {r^2} + {(h + b)^2}$ \cite{G}. Elements of energy-momentum tensor of the disk can be calculated using Einstein equations. With the help of  tetrads, one can calculate the values of energy density $\sigma = - T_t^t$, radial pressure $P_r = T_r^r$ which is equal to the azimuthal pressure $P_\varphi = T_\varphi ^\varphi $, and vertical pressure $P_z = T_z^z$, as well as the effective Newtonian density $\rho = \sigma + {P_r} + {P_z} + {P_\varphi}$  \cite{bib7}. 

\section{Thick disks in an expanding universe surrounded by dark energy}\label{section3}
In this part, by applying conformal transformation on static solution of thick disks in an expanding universe which is surrounded by dark energy, we solve Einstein's equations for thick gravitational disks.

Since the real world is not static and is accelerating, by applying a suitable conformal transformation on static metrics, one can switch to FRW spacetime as the expanding universe.

Technically, this can be done by obtaining a transformed metric $\overline g _{ab}$ from the original metric $g_{ab}$ as,
\begin{equation}
\ {\overline g _{ab}} = {\Omega ^2}({x^\alpha }){g_{ab}},
\end{equation}
where ${\Omega } ({x^\alpha})$ is a function of the coordinates and is referred to as a conformal factor. The flat FRW metric is expressed as,
\begin{equation}
\ d{s^2} = - d{t^2} + {R^2}(d{r^2} + {r^2}d{\Omega ^2}).
\end{equation}
By choosing $R(t)$ such that,
\begin{equation*}
\ dt = R(t)d{t_c},
\end{equation*}
we will find,
\begin{equation}
\ d{s^2} = {R^2}(t)\left[ { - d{t_c^2} + d{r^2} + {r^2}d{\Omega ^2}} \right].
\end{equation}
This is in fact the transformed Schwarzschild metric under conformal transformation with the conformal factor $R(t)$. We notice that under conformal transformations, null geodesics are still conserved.

In the next section, we will apply the conformal transformation on static metric of thick disks to find the metric of thick disks in an expanding universe.

	\subsection{The effect of the accelerating expansion of the universe on the static thick disk's solution}\label{section3.2}
	
As it is mentioned before, it is possible to obtain Schwarzschild metric in FRW spacetime by applying conformal transformations on Schwarzschild metric in static spacetime.

Also,  by using the "displace, cut, fill and reflect" method on the Schwarzschild  metric, we find the metric of a static thick disk. Further, by applying a conformal transformation on static metric, the metric of a thick disk is obtained in expanding spacetime. Actually, this solution will give the results more relevant to the real world.

In addition, by solving Einstein equations in the presence of cosmological constant and finding the energy-momentum tensor, one can compute other properties of the disk such as the energy density, mass density, together with the azimuthal, radial and vertical pressures.

As we know, isotropic metric in the cylindrical coordinate system is described by \eqref{eq2},\eqref{eq3}.
Utilizing the transformation $z \to h(z) + b$ described in section \ref{section2.2} we will obtain,
\begin{equation}\label{eq11}
\ \begin{array}{l}
\nu = \ln (\frac{{2\sqrt {{r^2} + {{(h + b)}^2}} - m}}{{2\sqrt {{r^2} + {{(h + b)}^2}} + m}}),\\
\mu = 2ln(1 + \frac{m}{{2\sqrt {{r^2} + {{(h + b)}^2}} }}).
\end{array}
\end{equation}

The continuity of the metric function and its derivatives on $z = \pm a$ is determined with the continuity of $h(z)$ and $h'(z)$ on $z = \pm a$. Equations \eqref{eq11} are the thick disk's metric in the static spacetime that we solved in Section \ref{section2}.

By applying conformal transformation on metric \eqref{eq11}, one can find the metric of thick disks in an expanding universe inthe form of \eqref{eq2}, with following variables,
\begin{equation}\label{eq13}
\begin{array}{l}
\nu = \ln \frac{{2Ra(t) - m}}{{2Ra(t) + m}},\\
\mu = 2ln(1 + \frac{m}{{2Ra(t)}}).
\end{array}
\end{equation}
Substituting $R = \sqrt {{r^2} + {{(h + b)}^2}}$, we will find,
\begin{equation}\label{eq14}
\ \begin{array}{l}
\nu = \ln (\frac{{2a(t)\sqrt {{r^2} + {{(h + b)}^2}} - m}}{{2a(t)\sqrt {{r^2} + {{(h + b)}^2}} + m}}),\\
\mu = 2lna^{1/2}(1 + \frac{m}{{2a(t)\sqrt {{r^2} + {{(h + b)}^2}} }}).
\end{array}
\end{equation}
Where $a(t)$ is scale factor, which will be studied in the the following section.
	
\subsection{Finding the evolution of scale factor $a(t)$ from the dark energy equation of state in accelerating expanding universe}\label{section3.3}
According to the experimental data of cosmology, our universe undergoes an accelereting expansion phase. For more realistic thick disk's solutions, this effect should be taken into account. As a model, the effect of this acceleration on Einstein equations is considered by adding the cosmological constant to Einstein equations and the final effect will be seen in function a(t) \cite{CM}, which will be discused in detail in the current section.

To calculate the temporal evolution of $a(t)$, we start with Einstein equations. The effect of dark energy in Einstein equations is implemented by adding a constant term, the cosmological constant $\Lambda$, to the initial Einstein equations. This equation, after such addition, can be expressed as,
\begin{equation}\label{eq10}
\ {R_{\mu \nu }} - \frac{1}{2}{g_{\mu \nu }}R - \Lambda {g_{\mu \nu }} = - 8\pi G{T_{\mu \nu }}.
\end{equation}
The relationship between the cosmological constant and energy density is as follows,
\begin{equation}
\ \Lambda = 8\pi G{\rho _D},
\end{equation}
where $\rho_D$ is the dark energy density and $ \Lambda = 2\times 10^{ - 35} m^{-2} $ \cite{bib19}.

Given $P = \varpi \rho$ as the equation of the state, P is pressure and $\rho$ is matter density , and using Einstein equations and FRW relations we will find \cite{bib17},
\begin{equation} \label{eq17}
\ {{\rm H}^2}  + \frac{\rm{K}}{a^2} = \frac{8 \pi G}{3}\rho + \frac{\Lambda}{3} ,
\end{equation}
\begin{equation} \label{eq18}
\ {{\rm H}^2}  + \dot {\rm H} = - \frac{4\pi G}{3}\rho (1 + 3\varpi ) + \frac{\Lambda }{3},
\end{equation}
\begin{equation}
 \dot \rho + 3(1 + \varpi )\rho {\rm H} = - \frac{{\dot \Lambda }}{{8\pi G}},
\end{equation}
where $H=\dot {\rm a}(t)/a(t)$ is the Hubble parameter and $K$ is the curvature constant which assumes values $-1$, $0$ and $+1$ for open, flat and close models of the universe .

It is noteworthy that to solve Einstein equations for relativistic thick disk in FRW spacetime, one needs to consider the metric in expanding spacetime and then solve Einstein equations in the presence of the cosmological constant. Hence, we need to compute the function $a(t)$ under the condition where $\Lambda \ne 0$. Here we have $\Lambda = 8\pi G{\rho _D}$ and assume $K=0$ which means flat universe, equations \eqref{eq17} and \eqref{eq18} can be written as,
\begin{equation} \label{eq6}
\ {{\rm H}^2} = \frac{{8\pi G}}{3}(\rho + {\rho _D}),
\end{equation}
\begin{equation} \label{eq7}
\ {{\rm H}^2} + \dot {\rm H} = - \frac{{4\pi G}}{3}\left[ {\rho (1 + 3\varpi ) + 2{\rho _D}} \right].
\end{equation}
To simplify the solutions, one can assume that the matter density is proportional to the dark energy density, $\rho_D = f\rho$. 
Here, we will find,
\begin{equation}
\ {{\rm H}^2} = \frac{{8\pi G}}{3}\rho (1 + f),
\end{equation}
\begin{equation}
\ {{\rm H}^2} + \dot {\rm H} = - \frac{{4\pi G}}{3}\rho \left[ {1 + 3\varpi + 2f} \right].
\end{equation}
By applying a similar calculation, we will have,
\begin{equation} \label{eq24}
\ a(t) = {a_0}{(t/t_0)^{\frac{{2(1 + f)}}{{3 + 3\varpi + 4 f}}}}.
\end{equation}

All of the effects introduced by $\Lambda \ne 0$ are contained in parameter $f$. In fact, the effect of a non-zero cosmological constant, led to the usage of $\Lambda = 8\pi G \rho_D$ and by making the assumption of $\rho_D = f{\rho}$, we arrived at the above result. In Equation \eqref{eq24}, $f$ denotes the effect of the cosmological constant. Further, a non-zero cosmological constant is indicative of the presence of acceleration, and acceleration is a result of the existence of dark energy. By substituting the obtained $a(t)$ function in the metric \eqref{eq14}, we solved Einstein equations in the presence of the cosmological constant and by finding $T_{\mu \nu}$, we calculated the values of ${\tilde \rho }$, ${\tilde \sigma }$, ${\tilde P}_r$, ${\tilde P}_{\varphi}$, and ${\tilde P}_z$.

\subsection{Exact solution to the Einstein equations for thick disks in FRW spacetime}\label{section3.4}
The effect of the acceleration of universe expansion appeared in the function a(t) in FRW spacetime of the form \eqref{eq24}. Hence, we will arrive to the final solution, by substituting ${a_0}$ , $f$ and $\varpi $ . Based on \cite{bib17} $\varpi $ shall be of,
\[ - 1.33 < \varpi < - 0.8\]
Although $ f $ changes by the time evolution, we have assumed that $f \simeq 15$ , since the variation rate is considerably slow and having a constant $f$ does not affect our calculations. Hence, considering that $f=\frac{{{\rho _{_D}}}}{\rho }$, we conclude that the world is surrounded by \%68.3 dark energy, \%4.9 ordinary matter and \%26.8 dark matter. Additionally, $a_0$ is a coefficient with the value of 1  \cite{bib18}, \cite{bib19}.

Next, from the Einstein equation \eqref{eq10}, we calculate the elements of energy-momentum tensor for thick disk in FRW spacetime. By choosing tetrads \cite{bib20}, such that,
\begin{equation}
\begin{array}{l}
{V^a} = {e^{ - \nu }}(1,0,0,0),\\
{W^a} = {e^{ - \mu }}(0,1,0,0),\\
{Y^a} = {e^{ - \mu }}(0,0,1,0),\\
{Z^a} = {e^{ - \mu }}(0,0,0,1).
\end{array}
\end{equation}
and by using equation \eqref{eq10}, we write energy-momentum tensor as we explained in section \ref{section2.2} as ,
\begin{equation}
 {T_{ab}} = \sigma {V_a}{V_b} + {P_r}{W_a}{W_b} + {P_z}{Y_a}{Y_{_b}} + {P_\varphi }{Z_a}{Z_b}.
\end{equation}
Also, By inserting the metric elements in the Einstein equation, we have found the energy-momentum tensor's elements and as a result, we have obtained the energy density and pressure components (appendix \ref{App A}). Following graphs are drawn for a thick disk with parameters $\tilde m = 1$ ,$\tilde b = 2$ ,$\tilde c= 1$ in a constant time and also, some levels curves of the density are displayed on the right graphs.

\begin{figure}[H]
\centering
 \includegraphics[scale=0.31]{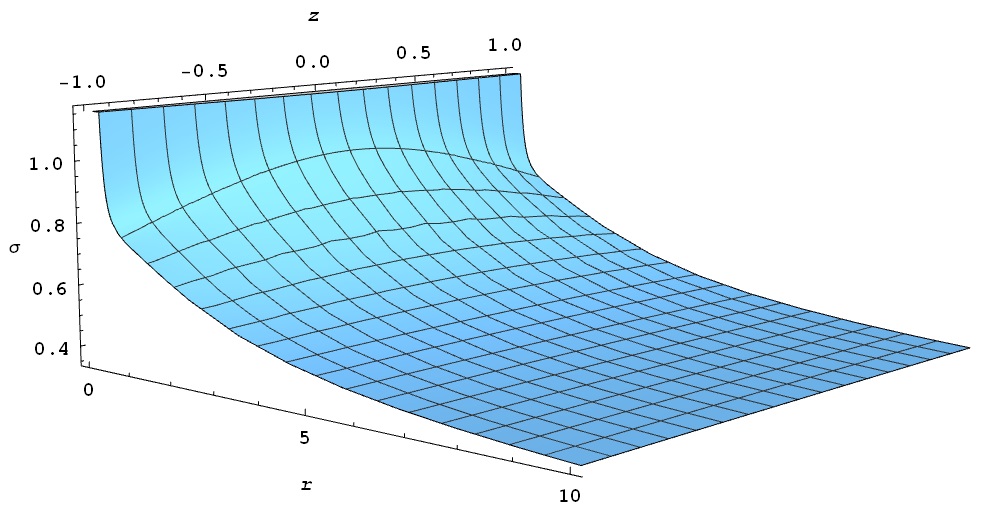}
\includegraphics[scale=0.295]{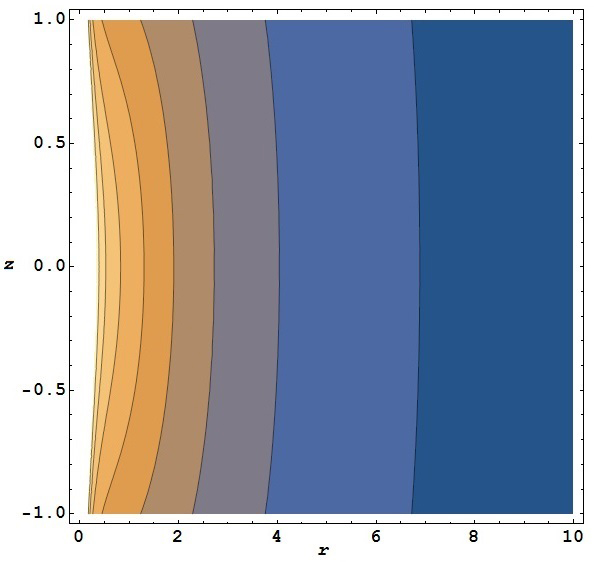}
 \caption{The energy density Eq. \eqref{g44} }
  \label{sigmat}
 \end{figure}

\begin{figure}[H]
\centering
\includegraphics[scale=0.35]{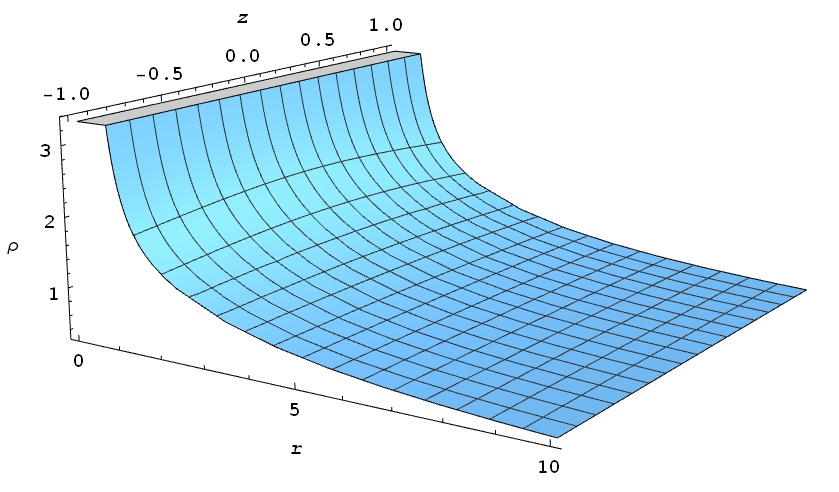}
\includegraphics[scale=0.30]{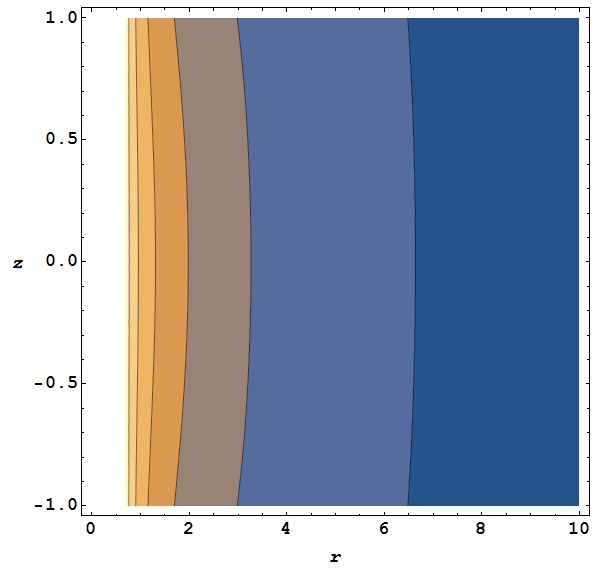}
\caption{The mass density Eq. \eqref{rho} for a thick disk with parameters ${{\tilde m}} = 1$ ,${{\tilde b}} = 2$ ,${{\tilde c}} = 1$ in a constant time. Some levels curves of the density are displayed on the right graph.}
\label{rhotsabet}
\end{figure}

\begin{figure}[H]
\centering
 \includegraphics[scale=0.32]{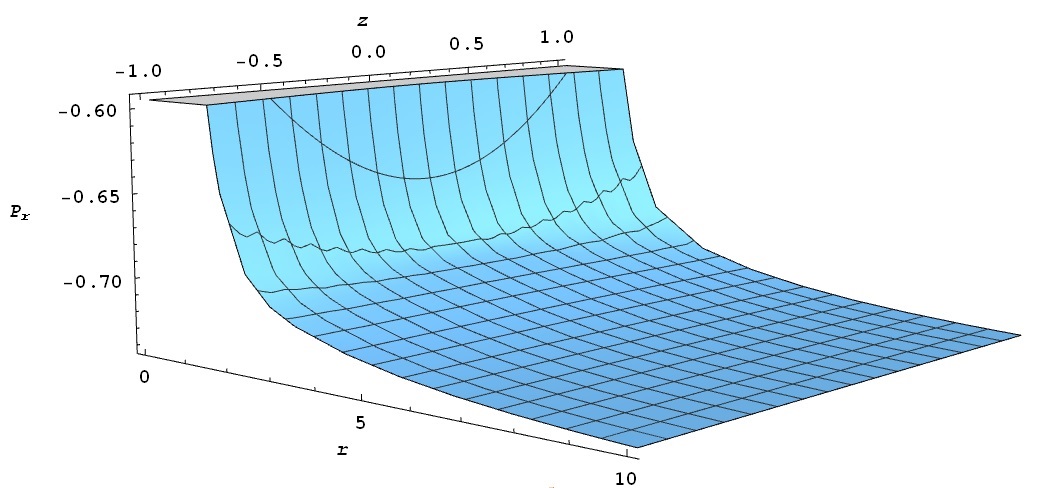}
 \includegraphics[scale=0.31]{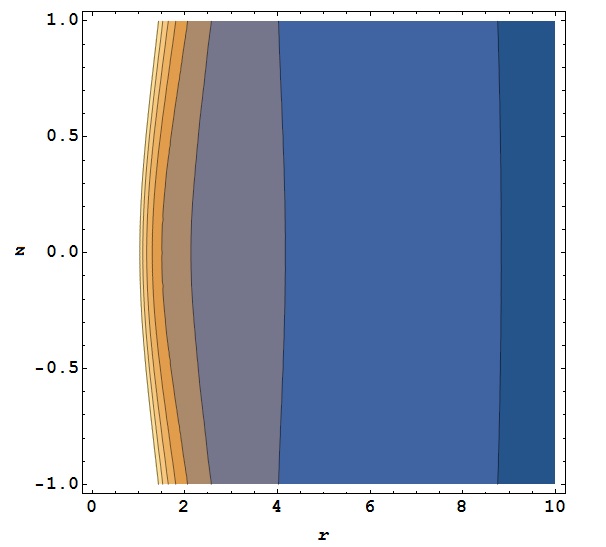}
 \caption{Radial pressures Eq. \eqref{g11} for a thick disk with parameters ${{\tilde m}} = 1$ ,${{\tilde b}} = 2$ ,${{\tilde c}} = 1$ in a constant time.}
  \label{Prtsabet}
 \end{figure}

\begin{figure}[H]
\centering
\includegraphics[scale=0.33]{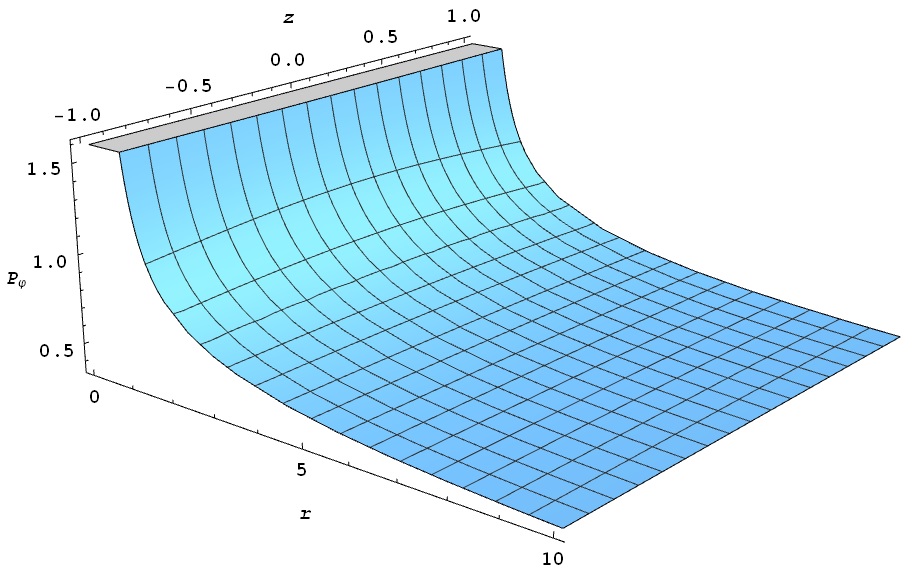}
\includegraphics[scale=0.31]{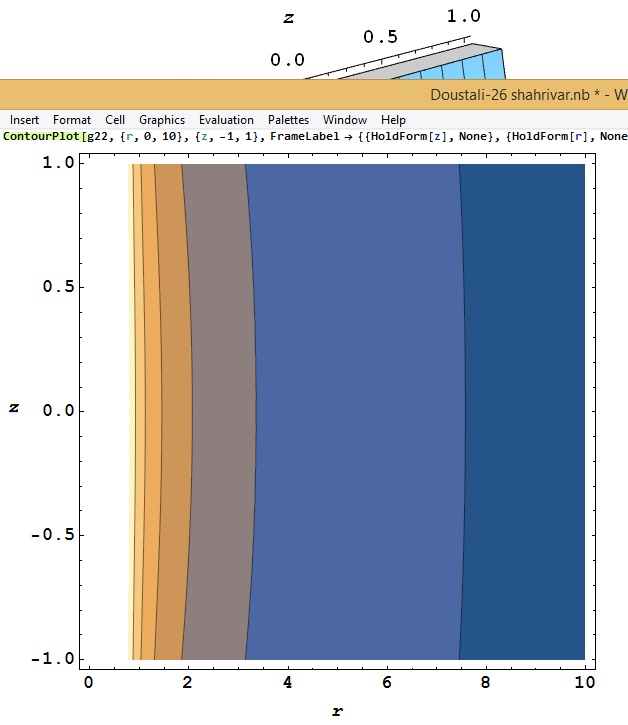}
\caption{Azimuthal pressure Eq. \eqref{g22} for a thick disk with parameters ${{\tilde m}} = 1$ ,${{\tilde b}} = 2$ ,${{\tilde c}} = 1$ in a constant time.}
\label{Pfitsabet}
\end{figure}

\begin{figure}[H]
\centering
\includegraphics[scale=0.37]{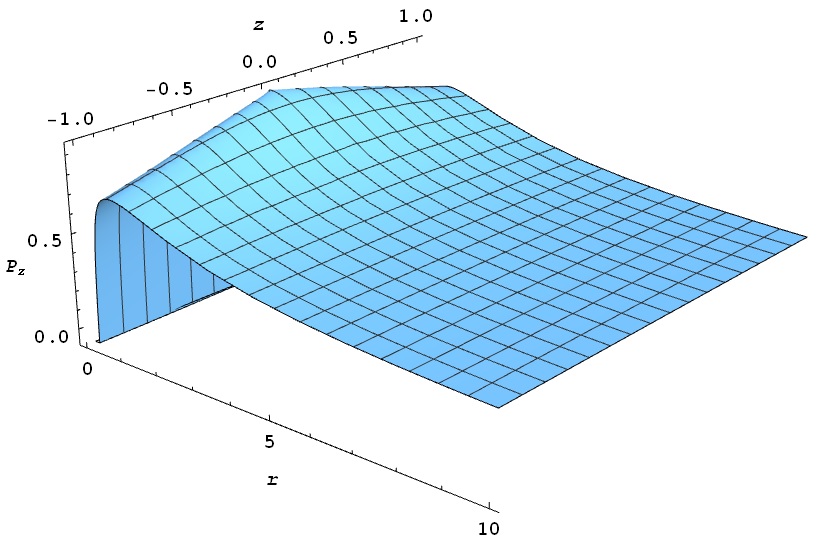}
\includegraphics[scale=0.31]{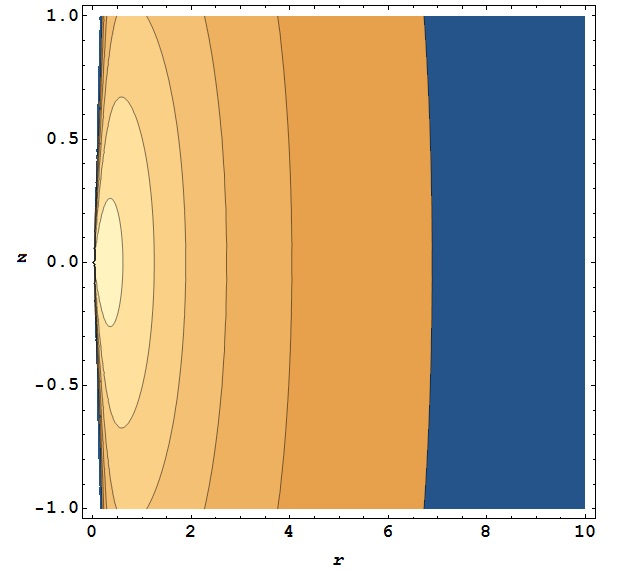}
\caption{Vertical pressure Eq. \eqref{g33} for a thick disk with parameters ${{\tilde m}} = 1$ ,${{\tilde b}} = 2$ ,${{\tilde c}} = 1$ in a constant time.}
\label{Pztsabet}
\end{figure}

In figures \ref{sigmat}, \ref{rhotsabet}, \ref{Prtsabet}, \ref{Pfitsabet}, \ref{Pztsabet} surfaces and levels curves of $\tilde \sigma$, $\tilde \rho$ , ${{\tilde P}_r}$, ${{\tilde P}_\varphi}$ and ${{\tilde P}_z}$ for schwartzschild thick disk in $FRW$ spacetime in a constant time are shown. These curves indicate certain features of the disk: for example they show that the density increases towards the disk's center.

\begin{figure}[H]
\centering
\includegraphics[scale=0.32]{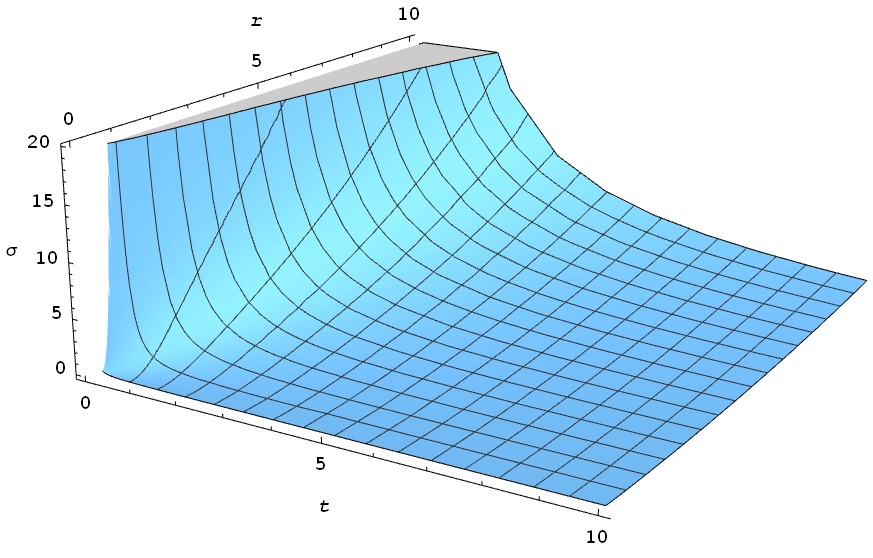}
\includegraphics[scale=0.335]{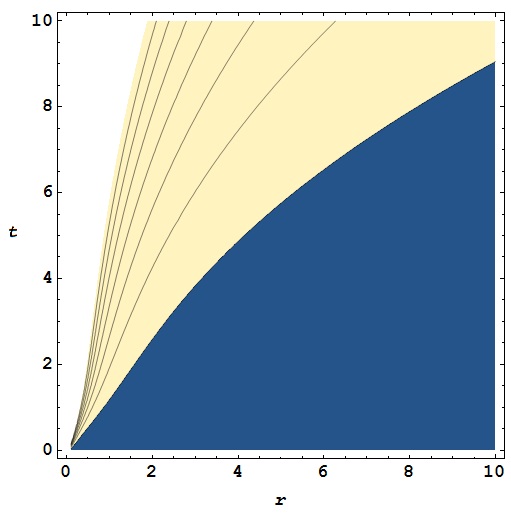}
\caption{Checking weak energy condition in  ${{\ z}} = 0$}
\label{weakz0}
\end{figure}

\begin{figure}[H]
\centering
\includegraphics[scale=0.33]{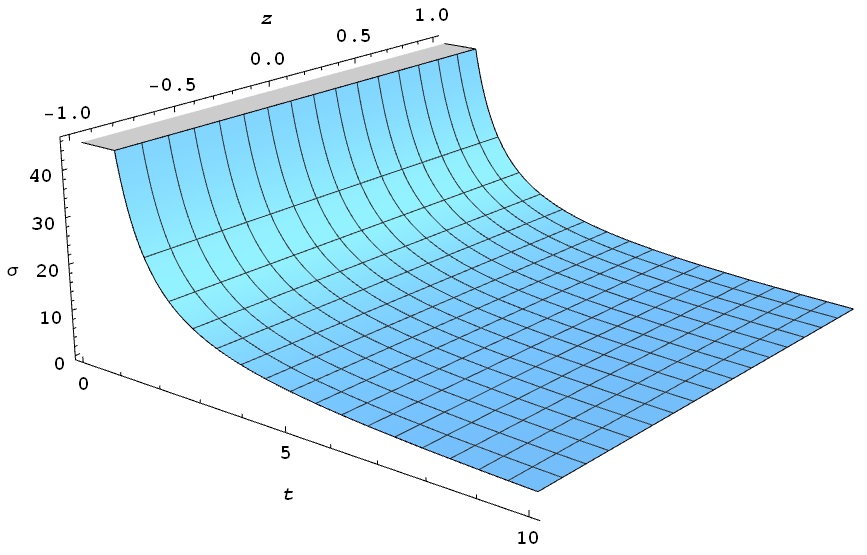}
\includegraphics[scale=0.31]{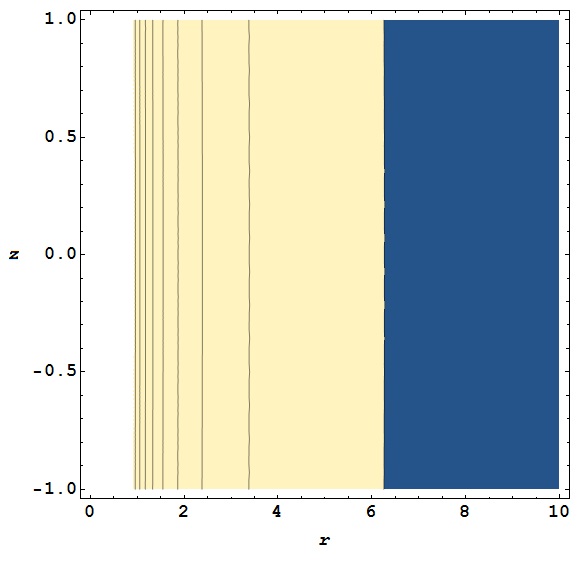}
\caption{Checking weak energy condition in ${{\ r}} = 10$}
\label{weakr10}
\end{figure}

\begin{figure}[H]
\centering
\includegraphics[scale=0.34]{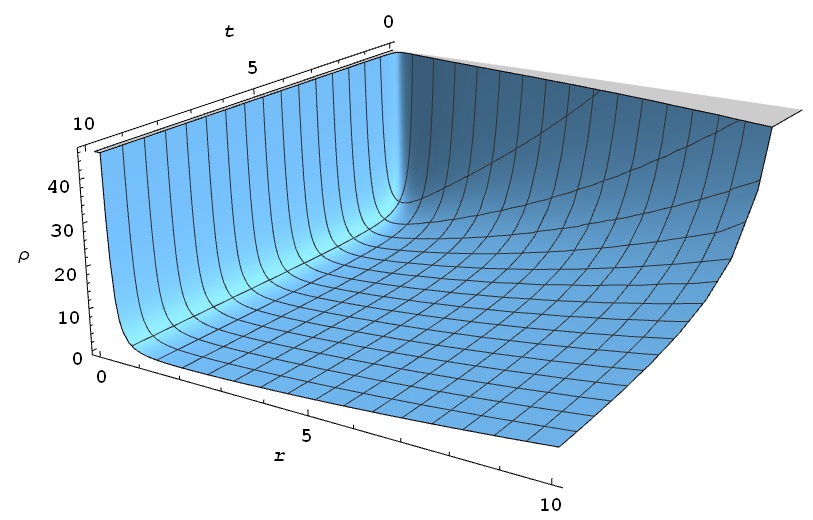}
\includegraphics[scale=0.33]{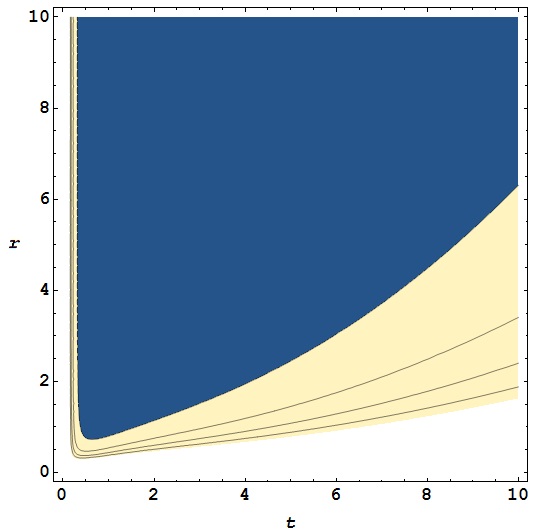}
\caption{Checking strong energy condition in ${{\ z}} = 0$ }
\label{Przrsabet4}
\end{figure}

\begin{figure}[H]
\centering
\includegraphics[scale=0.33]{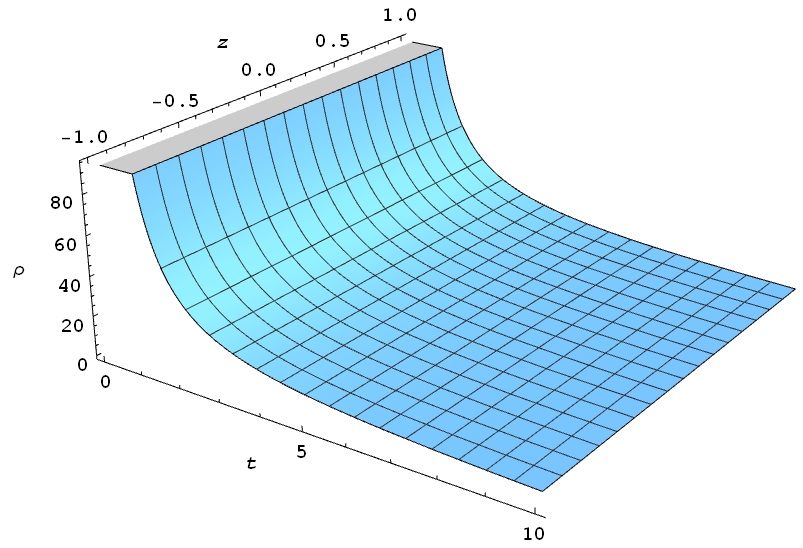}
\includegraphics[scale=0.32]{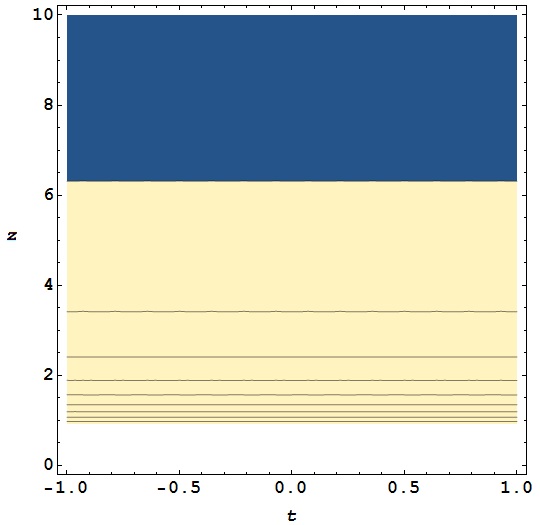}
\caption{Checking strong energy condition in ${{\ r}} = 10$ }
\label{Przrsabet5}
\end{figure}

\begin{figure}[H]
\centering
\includegraphics[scale=0.33]{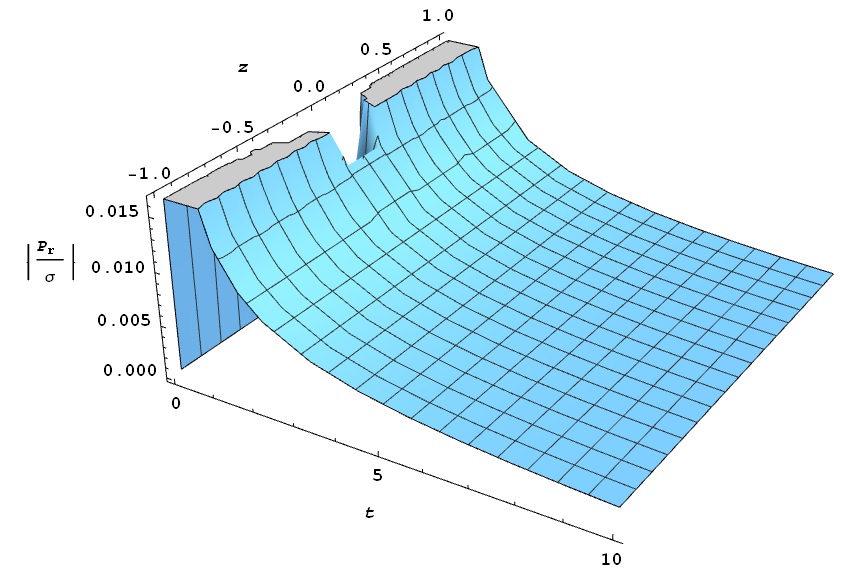}
\includegraphics[scale=0.3]{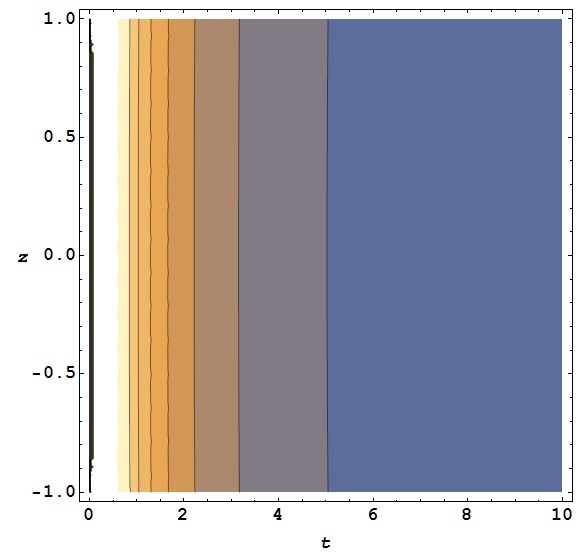}
\caption{Checking dominant energy condition  $ \vert\frac{p_{r}}{\sigma} \vert <1 $ in ${{\ r}} = 10$.}
\label{prsigma}
\end{figure}

\begin{figure}[H]
\centering
\includegraphics[scale=0.32]{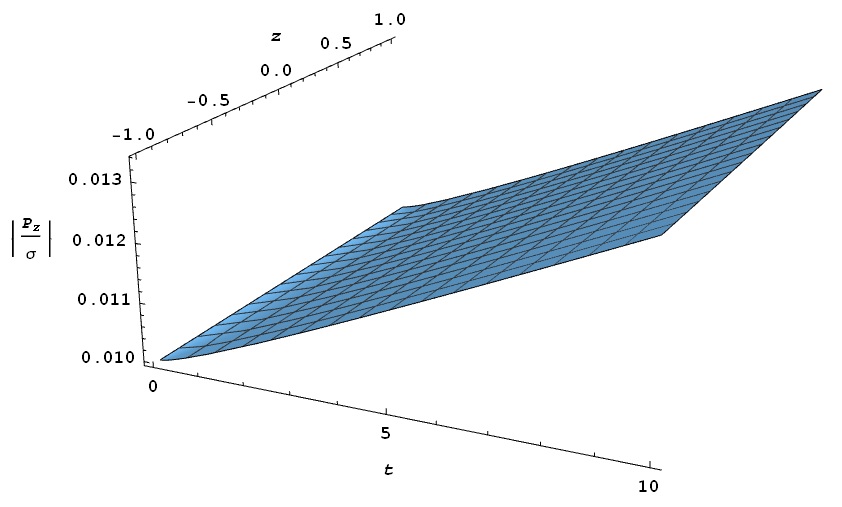}
\includegraphics[scale=0.31]{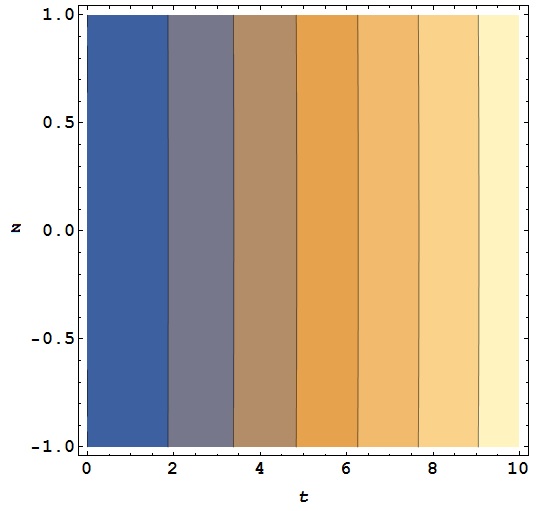}
\caption{Checking dominant energy condition  $ \vert\frac{p_{z}}{\sigma}\vert <1$ in ${{\ r}} = 10$.}
\label{pzsigma}
\end{figure}

\begin{figure}[H]
\centering
\includegraphics[scale=0.34]{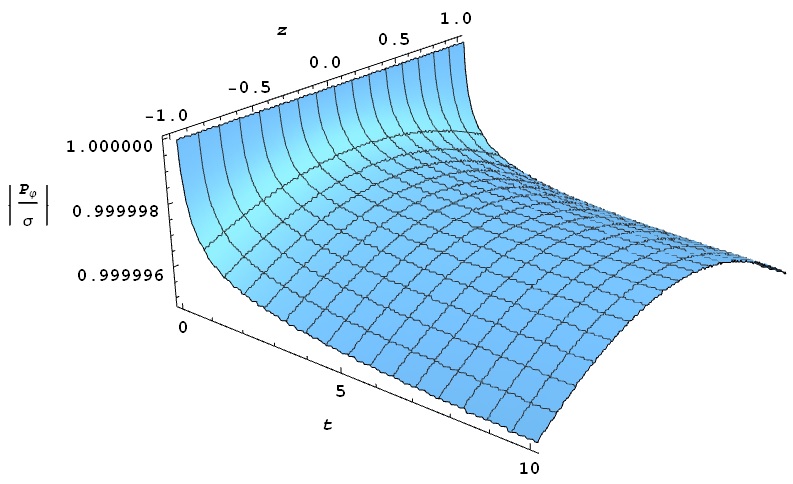}
\includegraphics[scale=0.3]{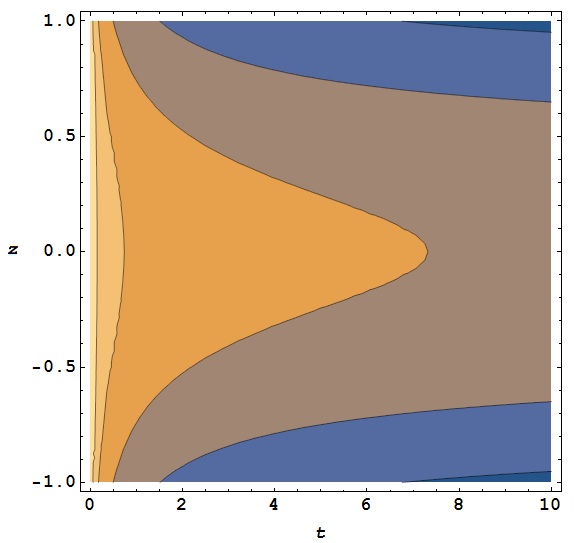}
\caption{Checking dominant energy condition $ \vert \frac{p_{\varphi}}{\sigma}\vert <1$ in ${{\ r}} = 10$.}
\label{pfisigma}
\end{figure}

Then we investigated the following dominated energy conditions,\\
I) weak energy condition $\sigma \ge 0$ ,\\
II) strong energy condition $ \rho = \sigma + {P_r} + P{}_\varphi + {P_z} \ge 0$ , \\
III) dominant energy conditions  $\left| {\frac{{{P_z}}}{\sigma }} \right|\le 1$ , $\left| {\frac{{{P{}_\varphi}}}{\sigma }} \right|\le 1$, $\left| {\frac{{{P_r}}}{\sigma }} \right|\le 1$, \\
Since analytical calculations were complicated , we checked the energy conditions by drawing charts which are shown in figures \ref{weakz0}, \ref{weakr10}, \ref{Przrsabet4}, \ref{Przrsabet5}, \ref{prsigma}, \ref{pzsigma} and \ref{pfisigma}. Finally we found that relativistic thick disks in FRW spacetime satisfy all above mentioned conditions.

\section{Conclusion}
We know that the real universe is in an accelerated expansion phase due to the existence of dark energy. Hence in this article, we modelled the effect of the accelerating expanding universes under the influence of dark energy on the galaxies with the help of thick disks.
 
To do such an investigation we reviewed "displace, cut, fill and reflect" method, to obtain the equations of thick disks from thin disks. Then, with help of the conformal transformations, the metric of the expanding thick disk from the static thin disk. Finally, by using the Einstein's equations and entering the density of dark energy in $\Lambda$ term, the evolution equation of $a(t)$ in the metric of the expanding thick disk is obtained.

Using the equations \eqref{eq13}, \eqref{eq14} and $a(t)$ obtained in \eqref{eq24}, and to have a deeper understanding of thick disks in FRW spacetime the corresponding energy-momentum tensor's elements and consequently the energy and mass density, radial, azimuthal, and vertical pressures are calculated in Appendix \ref{App A} and corresponding diagrams were demonstrated. Our results show that the energy and mass density, radial, azimuthal, and vertical pressures decreases by getting away from the center of the disc, in accordance with the expected physical properties and it shows spread out of disk which is expected. Also, we found that relativistic thick disks in FRW spacetime satisfy all the energy conditions.

Lastly, these results inspire some phenomenological researches for future works, such as investigating the effect of such a dark energy for another coordinated systems, or entering the effect of the dark energy in other terms more than the cosmological constant. Particularly, and in the case of doing an enough precise measurements on experimental data, such an investigations can pave the way for better understanding of the properties of the universe accelerating expansion mechanism, and also the properties of dark energy.

\appendix

\section{Appendix}\label{App A}
In this section the Einstein's equation's solutions are mentioned. Here، ${{\tilde P}_r} = {a^2}{P_r}$،, ${{\tilde P}_\varphi } = {a^2}{P_\varphi }$،, ${{\tilde P}_z} = {a^2}{P_z}$،, $\xi = \sqrt {4{r^2} + \psi } $،, $\psi = {({z^2} + 4)^2}$ and $a$ is equal to half of disk's thickness.
\begingroup\makeatletter\def\f@size{8}\check@mathfonts
\begin{eqnarray}
&& \tilde{\sigma} = \frac{1}{4 \xi  r^2 t^2  (\xi    \sqrt{t}+1 )^4  (\xi ^4 t-1 )}(64 r^8  (12 t^{5/2}+\xi  t^3 )+16 r^6 t^{3/2} (3 \xi  t^{3/2} \psi +45 \xi  \sqrt{t}+36 t \psi +80 )+4 r^4 \sqrt{t}  \nonumber \\
&& \qquad (3 \xi  t^{5/2} \psi ^2+90 \xi
    t^{3/2} \psi +36 t^2 \psi ^2+75 \xi  \sqrt{t}+160 t \psi +36 )+r^2 (7 \xi +12 t^{5/2} \psi ^3+80 t^{3/2} \psi ^2+\xi  t^3  \nonumber \\
&& \qquad (z^{12}+24 z^{10}+240 z^8+1280    z^6+3904 z^4+6848 z^2+5376 )+45 \xi  t^2 \psi ^2+75 \xi  t \psi +36 \sqrt{t} \psi )  \nonumber \\
&& \qquad -16 \xi  t^3 \psi  (z^2-4 )) \label{g44}\\
&&\tilde{P}_{r} = -\frac{1}{4 \xi  r^2 t^2 (\xi  \sqrt{t}-1)^2 (\xi  \sqrt{t}+1)^6}((\xi ^4 t-1) (192 r^8(4 t^{5/2}+\xi  t^3)+16 r^6 t^2 (15 \xi -16 \xi  t^2+9 \xi  t \psi +36 \sqrt{t} \psi ) \nonumber \\
&& \qquad -4 r^4 (-36    t^{5/2} \psi ^2-64 t^{7/2} (3 z^4+30 z^2+56)+48 \xi  t^4 \psi -\xi  t^3 (9 z^8+144 z^6+864 z^4+2304 z^2+2320) \nonumber \\
&& \qquad -30 \xi  t^2 \psi +15 \xi  t+12    \sqrt{t})+r^2 (-3 \xi +12 t^{5/2} \psi ^3+64 t^{7/2} \psi (3 z^4+24 z^2+64)-48 \xi  t^4 \psi ^2+\xi  t^3 (3 z^{12} \nonumber \\
&& \qquad +72 z^{10}+720 z^8+3840  z^6+11360 z^4+16768 z^2+9216)+15 \xi  t^2 \psi ^2-15 \xi  t \psi -12 \sqrt{t} \psi)-4 t^3 (\xi  t \psi ^3 \nonumber \\
&& \qquad +8 \sqrt{t} \psi ^2 (3 z^2-4)-\xi  \psi (z^4+32 z^2-16)))), \label{g11} \\
&&\tilde{P}_{\phi} = \frac{1}{4 r^2 t^4 (4 r^2+\psi )^{5/2} (\frac{1}{\xi  \sqrt{t}}+1)^4 (\xi ^2
   t-1)}(64 r^8 (12 t^{5/2}+\xi  t^3 )+16 r^6 t^{3/2} (3 \xi  t^{3/2} \psi +16 \xi  t^{5/2}    +45 \xi  \sqrt{t} \nonumber \\
&& \qquad+36 t \psi +80 )+4 r^4  (36 t^{5/2} \psi ^2+160    t^{3/2} \psi +48 \xi  t^4 \psi +\xi  t^3 (3 z^8+48 z^6+288 z^4+768 z^2+752 ) +75 \xi  t\nonumber \\
&& \qquad +90 \xi  t^2 \psi +36 \sqrt{t} )+r^2  (7 \xi +12 t^{5/2} \psi ^3+80   t^{3/2} \psi ^2+48 \xi  t^4 \psi ^2+\xi  t^3  (z^{12}+24 z^{10}+240 z^8+1280 z^6\nonumber \\
&& \qquad +3872 z^4+6592 z^2+4864 )+45 \xi  t^2 \psi ^2+75 \xi  t \psi +36 \sqrt{t} \psi  )+4 \xi  t^3  (t \psi ^3-\psi  z^2  (z^2+12 ) )) \label{g22} \\
&&\tilde{P}_{z} = \frac{}{4 \xi  r^2 t^2 (\xi  \sqrt{t}+1 )^4 (\xi ^4 t-1 )}(64 r^8  (12 t^{5/2}+\xi  t^3 )+16 r^6 t^{3/2}  (3 \xi  t^{3/2} \psi +45 \xi  \sqrt{t}+36 t \psi +80 )+4 r^4 \sqrt{t}  \nonumber \\
&& \qquad (3 \xi  t^{5/2} \psi ^2+90 \xi     t^{3/2} \psi +36 t^2 \psi ^2+75 \xi  \sqrt{t}+160 t \psi +36 )+r^2 (7 \xi +12 t^{5/2} \psi ^3+80 t^{3/2} \psi ^2+\xi  t^3 \psi   \nonumber \\
&& \qquad (z^8+16 z^6+96 z^4+256    z^2+320 )+45 \xi  t^2 \psi ^2+75 \xi  t \psi +36 \sqrt{t} \psi )-64 \xi  t^3 \psi  z^2) \label{g33}\\
&&\tilde{\rho} =\tilde{\sigma} + \tilde{P}_{r} + \tilde{P}_{\varphi} +\tilde{P}_{z} \label{rho}
\end{eqnarray}
\endgroup

\end{document}